\documentclass[12pt]{iopart}
\usepackage{iopams}
\begin{document}
\title[Modified strip projection method]{Modified strip projection method}
\author{Nicolae Cotfas}
\address{Faculty of Physics, University of Bucharest,
PO Box 76-54, Post Office 76, Bucharest, Romania, 
E-mail address: ncotfas@yahoo.com } 
\begin{abstract}
The diffraction image of a quasicrystal admits a finite group $G$ as a 
symmetry group, and the quasicrystal can be regarded as a quasiperiodic
packing of copies of a $G$-cluster $\mathcal{C}$, joined by glue atoms. 
The physical space $E$ containing $\mathcal{C}$ can be embedded into
a higher-dimensional space $\mathbb{R}^k$ such that, up to an inflation 
factor, $\mathcal{C}$ is the orthogonal projection of the set 
$\{ (\pm 1,0,...,0)$, $(0,\pm 1,0,...,0)$, ... $(0,...,0,\pm 1)\}$.
The projections of the points of $\mathbb{Z}^k$ lying in the strip 
$S\!=\!E\!+\![-1/2,1/2]^k\!\!+\!t$ obtained by 
shifting a hypercube $[-1/2,1/2]^k\!\!+\!t$ along $E$
is a quasiperiodic packing of partially occupied copies of $\mathcal{C}$,
but unfortunately, the occupation of clusters is very low.
In our modified strip projection method we firstly determine for each point 
$x\in \mathbb{Z}^k\cap S$ the number $n(x)$ of all the arithmetic neighbours
of $x$ lying in the strip $S$, and project the points of $\mathbb{Z}^k\cap S$ 
on $E$ in the decreasing order of the occupation number $n(x)$.
In the case when $n(x)$ represents more
than $p\%$ of all the points of the cluster $C$ we project all the 
arithmetic neighbours of $x$ (lying inside or outside $S$).
We choose $p$ such that to avoid the superposition of the fully
occupied clusters. The projection of a point $x$ with $n(x)$
less than $p\%$ of all the points of the cluster $C$ is added to the 
pattern only if it is not too close to the already obtained points.
\end{abstract}
\maketitle

\section{Introduction}

Quasicrystals are materials with perfect long-range order, 
but with no three-dimensional translational periodicity.
The discovery of these solids in the early 1980's and the challenge 
to describe their structure led to a great interest in quasiperiodic 
sets of points \cite{KP,M}. The diffraction image of a quasicrystal
contains  a set of sharp Bragg peaks invariant under a finite
non-crystallographic group of symmetries $G$, called the symmetry group 
of quasicrystal (in reciprocal space). 
In the case of quasicrystals with no translational periodicity this group
is the icosahedral group $Y$ and in the case of quasicrystals 
periodic along one direction (two-dimensional quasicrystals) $G$ is one 
of the cyclic groups $C_8$ (octagonal quasicrystals), $C_{10}$ 
(decagonal quasicrystals) and $C_{12}$ (dodecagonal quasicrystals).

Real structure information obtained by high resolution transmission 
electron microscopy suggests us that a quasicrystal with symmetry group
$G$ can be regarded as a quasiperiodic packing of copies of 
a well-defined $G$-invariant cluster $\mathcal{C}$, joined by glue atoms \cite{J}. 
From a mathematical point of view, a $G$-cluster  is a finite union 
of orbits of $G$, in a fixed linear representation of $G$. A mathematical algorithm 
for generating quasiperiodic packings of interpenetrating copies of G-clusters
was obtained by author in collaboration with Verger-Gaugry several years ago \cite{C1}. 
This algorithm based on strip projection method \cite{E,KKL,K,KN} works for any finite group $G$
and any $G$-cluster, but in the case of a multi-shell cluster the dimension of the 
involved superspace is rather high, and the occupation of the clusters
occurring in the generated pattern is too low.
Some mathematical results recently obtained by author \cite{C3,C5} simplify the 
computer program and allow to use strip projection method in the superspaces
required by this approach. Now, our aim is to present a way to increase the
occupation of clusters occurring in the generated quasiperiodic set.

\section{Two-dimensional packings of clusters}

Let $G$ be one of the cyclic groups $C_8$, $C_{10}$, $C_{12}$.
Each group $C_n$ can be defined as
\begin{equation}
 C_n=\langle \ a \ |\ \ a^n=e\ \rangle =\{ \, e,\, a,\, a^2,\, ...,\, a^{n-1} \}
\end{equation}
and the formula
\begin{equation}
 a(\alpha ,\beta )=
\left(\alpha \, \cos \frac{2\pi }{n}-\beta \, \sin \frac{2\pi }{n}, \
      \alpha \, \sin \frac{2\pi }{n}+\beta \, \cos \frac{2\pi }{n} \right) 
\end{equation}
define an $\mathbb{R}$-irreducible representation in $\mathbb{R}^2$. 
The orbit generated by $(\alpha ,\beta )\not=(0,0)$
\begin{equation}
C_n(\alpha ,\beta )=\{ (\alpha ,\beta ), a(\alpha ,\beta ),\, 
a^2(\alpha ,\beta ),\, ...,\, a^{n-1}(\alpha ,\beta )\} 
\end{equation}
contains $n$ points (vertices of a regular polygon with $n$ sides). 
Let 
\begin{equation} 
\mathcal{C}_2=\{ v_1,\, v_2,\, ...,\, v_k,\, -v_1,\, -v_2,\, ...,\, -v_k \}
\end{equation} 
where $v_1=(v_{11},v_{21})$, $v_2=(v_{12},v_{22})$,..., $v_k=(v_{1k},v_{2k})$,
be a fixed $G$-cluster, that is, a finite union of orbits of $G$.
From the general theory \cite{C1} (a direct verification is also possible) it
follows that the vectors 
\begin{equation} 
w_1=(v_{11},v_{12},...,v_{1k})\qquad {\rm and}\qquad 
 w_2=(v_{21},v_{22},...,v_{2k})
\end{equation}
from $\mathbb{R}^{k}$ are orthogonal and have the same norm
\begin{equation} 
\begin{array}{l}
\langle w_1,w_2\rangle =v_{11}v_{21}+v_{12}v_{22}+...+v_{1k}v_{2k}=0\\[2mm] 
||w_1||=\sqrt{v_{11}^2+v_{12}^2+...+v_{1k}^2}=||w_2||.
\end{array} 
\end{equation}
We identify the physical space with the two-dimensional subspace
\begin{equation} 
{\bf E}_2=\{ \ \alpha w_1+\beta w_2 \ | \ \alpha,\, \beta \in \mathbb{R}\ \} 
\end{equation}
of the superspace $\mathbb{R}^k$ and denote by ${\bf E}_2^\perp $ the 
orthogonal complement 
\begin{equation} {\bf E}_2^\perp =\{ \ x\in \mathbb{R}^k\ |\ 
\langle x,y\rangle =0\ {\rm for\ all}\ y\in {\bf E}_2\ \}. \end{equation}
The orthogonal projection on ${\bf E}_2$ of a vector $x\in \mathbb{R}^k$ is the vector
\begin{equation} \label{pi}
\pi _2\, x= \left\langle x,\frac{w_1}{\kappa }\right\rangle\frac{w_1}{\kappa }+
             \left\langle x,\frac{w_2}{\kappa }\right\rangle\frac{w_2}{\kappa} 
\end{equation}
where $\kappa =||w_1||=||w_2||$, and the orthogonal projector corresponding 
to ${\bf E}_2^\perp $ is
\begin{equation} 
\pi _2^\perp :\mathbb{R}^{k}\longrightarrow {\bf E}_2^\perp \qquad
\pi _2^\perp x=x-\pi _2\, x. 
\end{equation}
We describe ${\bf E}_2$ by using the orthogonal basis 
$\{ \kappa ^{-2}w_1,\, \kappa ^{-2}w_2\}$. Therefore, in view of (\ref{pi}) the expression 
in coordinates of $\pi _2$ is
\begin{equation} 
\pi _2: \mathbb{R}^{k}\longrightarrow \mathbb{R}^2\qquad 
\pi _2x=(\langle x,w_1\rangle , \langle x,w_2\rangle ). 
\end{equation}

The projection ${\bf W}_{2,k}=\pi _2^\perp ({\Lambda }_{k})$ of the unit hypercube
\begin{equation}
{\Lambda }_{k}=
\left\{ (x_1,x_2,...,x_{k})\ \left|\ -\frac{1}{2}\leq x_i\leq \frac{1}{2}\ \
{\rm for\ all\ } i\in \{ 1,2,..., k\}\ \right. \right\} 
\end{equation}
is a polyhedron (called the {\it window of selection}) 
in the $(k\!-\!2)$-dimensional subspace ${\bf E}_2^\perp $, 
and each $(k\!-\!3)$-dimensional face of ${\bf W}_{2,k}$ is the projection of a 
$(k\!-\!3)$-dimensional face of ${\Lambda }_{k}$.
The vectors $e_1=(1,0,0,...,0)$, $e_2=(0,1,0,...,0)$, ..., $e_{k}=(0,0,...,0,1)$ from $\mathbb{R}^k$
form the canonical basis of $\mathbb{R}^{k}$, and each $(k\!-\!3)$-face of ${\Lambda }_{k}$ 
is parallel to $(k\!-\!3)$ of these vectors and orthogonal to three of them.
There exist eight $(k\!-\!3)$-faces of ${\Lambda }_{k}$ 
orthogonal to the distinct vectors $e_{i_1}$, $e_{i_2}$, $e_{i_3}$, and the set 
\begin{equation}\left\{ \ x=(x_1,x_2,...,x_{k})\ \left| \ \begin{array}{lcl}
x_i\in \{ -1/2,\, 1/2\} & {\rm if}&   i\in \{ i_1,\, i_2,\, i_3\} \\
x_i=0 & {\rm if}& i\not\in \{ i_1,\, i_2,\, i_3\}
\end{array} \right. \right\} \end{equation}
contains one and only one point from each of them.
There are 
\begin{equation} \left( \begin{array}{c}
k\\
3
\end{array}\right) =\frac{k(k-1)(k-2)}{6} \end{equation}
sets of $2^3$ parallel $(k\!-\!3)$-faces of ${\Lambda }_{k}$, and we label them by using the 
elements of the set
\begin{equation} \fl 
\mathcal{I}_{2,k}=\{ (i_1,i_2,i_3)\in \mathbb{Z}^3\ |\ 
1\leq i_1\leq k\!-\!2,\ \ i_1+1\leq i_2\leq k\!-\!1,\ \ i_2+1\leq i_3\leq k\ \}. 
\end{equation}

In $\mathbb{R}^3$ the cross-product of two vectors ${\bf v}=(v_x,v_y,v_z)$ and
${\bf w}=(w_x,w_y,w_z)$ is a vector orthogonal to ${\bf v}$ and ${\bf w}$, and
can be obtained by expanding the formal determinant
\begin{equation} {\bf v}\times {\bf w} = \left|
\begin{array}{ccc}
{\bf i} & {\bf j} & {\bf k} \\
v_x & v_y & v_z\\
w_x & w_y & w_z
\end{array} \right| \end{equation}
where $\{ {\bf i},\, {\bf j},\, {\bf k}\}$ is the canonical basis of $\mathbb{R}^3$. 
For any vector 
${\bf u}=(u_x,u_y,u_z)$, the scalar product of ${\bf u}$ and ${\bf v}\times {\bf w}$ is 
\begin{equation} {\bf u}({\bf v}\times {\bf w}) = \left|
\begin{array}{ccc}
u_x & u_y & u_z \\
v_x & v_y & v_z\\
w_x & w_y & w_z
\end{array} \right|. \end{equation}

In a very similar way, a vector $y$ orthogonal to $l=k\!-\!1$ vectors 
\begin{equation} u_i=(u_{i1},u_{i2},u_{i3},...,u_{ik})\qquad i\in \{ 1,2,3,...,l\} \end{equation} 
from $\mathbb{R}^{k}$ can be obtained by expanding the formal determinant
\begin{equation} y=\left| \begin{array}{ccccc}
e_1 & e_2 & e_3 & ... & e_{k}\\
u_{11} & u_{12} & u_{13} & ... & u_{1k}\\
u_{21} & u_{22} & u_{23} & ... & u_{2k}\\
... & ... & ... & ... & ...\\
u_{l1} & u_{l2} & u_{l3} & ... & u_{lk}\
\end{array} \right| \end{equation}
containing the vectors of the canonical basis in the first row.
For any $x=(x_1,x_2,...,x_k)\in \mathbb{R}^{k}$, the scalar product of $x$ and $y$ is
\begin{equation} \langle x,y\rangle =\left| \begin{array}{ccccc}
x_1 & x_2 & x_3 & ... & x_{k}\\
u_{11} & u_{12} & u_{13} & ... & u_{1k}\\
u_{21} & u_{22} & u_{23} & ... & u_{2k}\\
... & ... & ... & ... & ...\\
u_{l1} & u_{l2} & u_{l3} & ... & u_{lk}\
\end{array} \right|. \end{equation}

For example, 
\begin{equation}  \fl
y=\left| \begin{array}{cccccccc}
e_1 & e_2 & e_3 & e_4 & e_5 & e_6 & ... & e_k\\
0 & 0 & 0 & 1 & 0 & 0 & \dots & 0\\ 
0 & 0 & 0 & 0 & 1 & 0 & ... & 0\\ 
0 & 0 & 0 & 0 & 0 & 1 & ... & 0\\ 
... & ... & ... & ... & ... & ... & ... & ...\\ 
0 & 0 & 0 & 0 & 0 & 0 & ... & 1\\
v_{11} & v_{12} & v_{13} & v_{14} & v_{15} & v_{16} & ... & v_{1k}\\
v_{21} & v_{22} & v_{23} & v_{24} & v_{25} & v_{26} & ... & v_{2k}\\
\end{array} \right| 
= (-1)^{k-1}\left| \begin{array}{ccc}
e_1 & e_2 & e_3\\
v_{11} & v_{12} & v_{13}\\
v_{21} & v_{22} & v_{23}
\end{array} \right| \end{equation}
is a vector orthogonal to the vectors $e_4$, $e_5$, ...,  $e_{k}$, $w_1$, $w_2$, and
\begin{equation} \langle x,y\rangle =(-1)^{k-1}
\left| \begin{array}{ccc}
x_1& x_2 & x_3\\
v_{11} & v_{12} & v_{13}\\
v_{21} & v_{22} & v_{23}
\end{array} \right| \end{equation}
for any $x\in \mathbb{R}^{k}$. The vector  $y$ belongs to ${\bf E}_2^\perp $, and since
$e_i-\pi _2^\perp e_i$ is a linear 
combination of $w_1$ and $w_2$, it is also orthogonal to $\pi _2^\perp e_4$, 
$\pi _2^\perp e_5$, ..., $\pi _2^\perp e_{k}$. Therefore, $y$ is orthogonal 
to the $(k\!-\!3)$-faces of ${\bf W}_{2,k}$ labelled by $(1,2,3)$. Similar results can be 
obtained for any $(i_1,i_2,i_3)\in \mathcal{I}_{2,k}$.

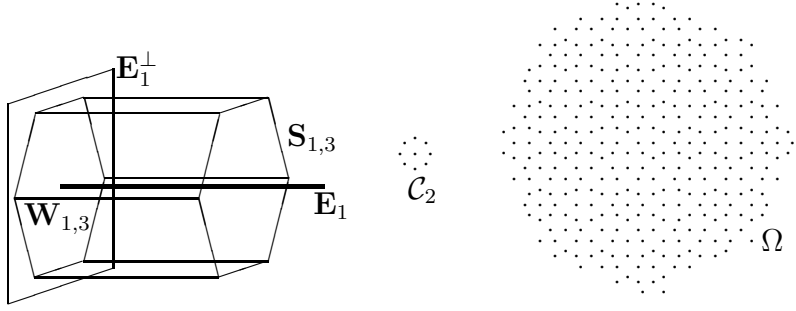
\begin{figure}
\setlength{\unitlength}{0.7mm}
\begin{picture}(130,60)(-30,0)
\put(62,17){${\bf E}_1$}
\put(80,20){$\mathcal{C}_2$}
\put(147,10){$\Omega $}
\put(24.5,43){${\bf E}_1^\perp $}
\put(7,15){${\bf W}_{1,3}$}
\put(57,30){${\bf S}_{1,3}$}
\put(4,0){\line(0,1){38}}
\put(4,0){\line(3,1){20}}
\put(4,38){\line(3,1){20}}
\put(24,6.8){\line(0,1){38}}
\put(9,5){\line(3,1){9.1}}
\put(9,5){\line(1,0){35}}
\put(9,5){\line(-1,4){3.8}}
\put(5.2,20){\line(1,4){4}}
\put(5.2,20){\line(1,0){35}}
\put(18.4,8){\line(1,4){4}}
\put(18.4,8){\line(1,0){35}}
\put(22.3,23.9){\line(-1,4){3.8}}
\put(22.3,23.9){\line(1,0){35}}
\put(9.2,36.3){\line(3,1){9.1}}
\put(9.2,36.3){\line(1,0){35}}
\put(18.4,39.1){\line(1,0){35}}
\put(44,5){\line(3,1){9.1}}
\put(44,5){\line(-1,4){3.8}}
\put(40.2,20){\line(1,4){4}}
\put(53.4,8){\line(1,4){4}}
\put(57.3,23.9){\line(-1,4){3.8}}
\put(44.2,36.3){\line(3,1){9.1}}
\linethickness{0.4mm}
\put(14,22.3){\line(1,0){50}}
\setlength{\unitlength}{2mm}
\put(   44.30000,  10.72426){\circle*{0.2}} 
\put(   43.30000,  10.72426){\circle*{0.2}} 
\put(   43.59289,  10.01716){\circle*{0.2}} 
\put(   45.00711,  11.43137){\circle*{0.2}} 
\put(   44.30000,   9.72426){\circle*{0.2}} 
\put(   44.30000,  11.72426){\circle*{0.2}} 
\put(   45.00711,  10.01716){\circle*{0.2}} 
\put(   42.59289,  10.01716){\circle*{0.2}} 
\put(   43.30000,  11.72426){\circle*{0.2}} 
\put(   42.59289,  11.43137){\circle*{0.2}} 
\put(   43.59289,   9.01716){\circle*{0.2}} 
\put(   46.00711,  11.43137){\circle*{0.2}} 
\put(   45.00711,  12.43137){\circle*{0.2}} 
\put(   45.71421,  10.72426){\circle*{0.2}} 
\put(   45.00711,   9.01716){\circle*{0.2}} 
\put(   43.59289,  12.43137){\circle*{0.2}} 
\put(   46.00711,  10.01716){\circle*{0.2}} 
\put(   41.59289,  10.01716){\circle*{0.2}} 
\put(   42.59289,   9.01716){\circle*{0.2}} 
\put(   41.88579,  10.72426){\circle*{0.2}} 
\put(   42.59289,  12.43137){\circle*{0.2}} 
\put(   44.30000,   8.31005){\circle*{0.2}} 
\put(   46.00711,  12.43137){\circle*{0.2}} 
\put(   46.71421,  10.72426){\circle*{0.2}} 
\put(   45.71421,  13.13848){\circle*{0.2}} 
\put(   45.00711,  13.43137){\circle*{0.2}} 
\put(   44.30000,  13.13848){\circle*{0.2}} 
\put(   46.00711,   9.01716){\circle*{0.2}} 
\put(   45.71421,   8.31005){\circle*{0.2}} 
\put(   41.59289,   9.01716){\circle*{0.2}} 
\put(   40.88579,  10.72426){\circle*{0.2}} 
\put(   41.88579,   8.31005){\circle*{0.2}} 
\put(   43.30000,   8.31005){\circle*{0.2}} 
\put(   41.88579,  11.72426){\circle*{0.2}} 
\put(   41.59289,  12.43137){\circle*{0.2}} 
\put(   43.30000,  13.13848){\circle*{0.2}} 
\put(   42.59289,  13.43137){\circle*{0.2}} 
\put(   41.88579,  13.13848){\circle*{0.2}} 
\put(   44.30000,   7.31005){\circle*{0.2}} 
\put(   45.00711,   7.60294){\circle*{0.2}} 
\put(   46.71421,  13.13848){\circle*{0.2}} 
\put(   46.71421,  11.72426){\circle*{0.2}} 
\put(   47.71421,  10.72426){\circle*{0.2}} 
\put(   47.42132,  11.43137){\circle*{0.2}} 
\put(   46.71421,   9.72426){\circle*{0.2}} 
\put(   47.42132,  10.01716){\circle*{0.2}} 
\put(   45.71421,  14.13848){\circle*{0.2}} 
\put(   44.30000,  14.13848){\circle*{0.2}} 
\put(   46.71421,   8.31005){\circle*{0.2}} 
\put(   40.88579,   8.31005){\circle*{0.2}} 
\put(   40.88579,   9.72426){\circle*{0.2}} 
\put(   40.17868,  10.01716){\circle*{0.2}} 
\put(   40.88579,  11.72426){\circle*{0.2}} 
\put(   40.17868,  11.43137){\circle*{0.2}} 
\put(   41.88579,   7.31005){\circle*{0.2}} 
\put(   42.59289,   7.60294){\circle*{0.2}} 
\put(   43.30000,   7.31005){\circle*{0.2}} 
\put(   40.88579,  13.13848){\circle*{0.2}} 
\put(   43.30000,  14.13848){\circle*{0.2}} 
\put(   41.88579,  14.13848){\circle*{0.2}} 
\put(   43.59289,   6.60294){\circle*{0.2}} 
\put(   45.00711,   6.60294){\circle*{0.2}} 
\put(   46.00711,   7.60294){\circle*{0.2}} 
\put(   47.71421,  13.13848){\circle*{0.2}} 
\put(   46.71421,  14.13848){\circle*{0.2}} 
\put(   47.42132,  12.43137){\circle*{0.2}} 
\put(   48.42132,  11.43137){\circle*{0.2}} 
\put(   48.42132,  10.01716){\circle*{0.2}} 
\put(   47.42132,   9.01716){\circle*{0.2}} 
\put(   45.00711,  14.84558){\circle*{0.2}} 
\put(   43.59289,  14.84558){\circle*{0.2}} 
\put(   47.71421,   8.31005){\circle*{0.2}} 
\put(   46.71421,   7.31005){\circle*{0.2}} 
\put(   40.17868,   7.60294){\circle*{0.2}} 
\put(   40.88579,   7.31005){\circle*{0.2}} 
\put(   40.17868,   9.01716){\circle*{0.2}} 
\put(   39.17868,  10.01716){\circle*{0.2}} 
\put(   39.47157,  10.72426){\circle*{0.2}} 
\put(   40.17868,  12.43137){\circle*{0.2}} 
\put(   39.17868,  11.43137){\circle*{0.2}} 
\put(   42.59289,   6.60294){\circle*{0.2}} 
\put(   40.88579,  14.13848){\circle*{0.2}} 
\put(   42.59289,  14.84558){\circle*{0.2}} 
\put(   44.30000,   5.89584){\circle*{0.2}} 
\put(   46.00711,   6.60294){\circle*{0.2}} 
\put(   45.00711,   5.60294){\circle*{0.2}} 
\put(   47.71421,  14.13848){\circle*{0.2}} 
\put(   48.42132,  12.43137){\circle*{0.2}} 
\put(   47.42132,  14.84558){\circle*{0.2}} 
\put(   46.00711,  14.84558){\circle*{0.2}} 
\put(   49.12843,  10.72426){\circle*{0.2}} 
\put(   48.42132,   9.01716){\circle*{0.2}} 
\put(   45.00711,  15.84558){\circle*{0.2}} 
\put(   44.30000,  15.55269){\circle*{0.2}} 
\put(   47.71421,   7.31005){\circle*{0.2}} 
\put(   48.42132,   7.60294){\circle*{0.2}} 
\put(   47.42132,   6.60294){\circle*{0.2}} 
\put(   39.17868,   7.60294){\circle*{0.2}} 
\put(   40.17868,   6.60294){\circle*{0.2}} 
\put(   39.47157,   8.31005){\circle*{0.2}} 
\put(   41.59289,   6.60294){\circle*{0.2}} 
\put(   39.17868,   9.01716){\circle*{0.2}} 
\put(   38.47157,  10.72426){\circle*{0.2}} 
\put(   39.17868,  12.43137){\circle*{0.2}} 
\put(   40.17868,  13.43137){\circle*{0.2}} 
\put(   39.47157,  13.13848){\circle*{0.2}} 
\put(   41.88579,   5.89584){\circle*{0.2}} 
\put(   42.59289,   5.60294){\circle*{0.2}} 
\put(   43.30000,   5.89584){\circle*{0.2}} 
\put(   41.59289,  14.84558){\circle*{0.2}} 
\put(   40.17868,  14.84558){\circle*{0.2}} 
\put(   43.30000,  15.55269){\circle*{0.2}} 
\put(   42.59289,  15.84558){\circle*{0.2}} 
\put(   41.88579,  15.55269){\circle*{0.2}} 
\put(   44.30000,   4.89584){\circle*{0.2}} 
\put(   46.00711,   5.60294){\circle*{0.2}} 
\put(   46.71421,   5.89584){\circle*{0.2}} 
\put(   45.71421,   4.89584){\circle*{0.2}} 
\put(   48.42132,  14.84558){\circle*{0.2}} 
\put(   48.42132,  13.43137){\circle*{0.2}} 
\put(   49.42132,  12.43137){\circle*{0.2}} 
\put(   49.12843,  13.13848){\circle*{0.2}} 
\put(   49.12843,  11.72426){\circle*{0.2}} 
\put(   47.42132,  15.84558){\circle*{0.2}} 
\put(   46.71421,  15.55269){\circle*{0.2}} 
\put(   46.00711,  15.84558){\circle*{0.2}} 
\put(   50.12843,  10.72426){\circle*{0.2}} 
\put(   49.12843,   9.72426){\circle*{0.2}} 
\put(   49.42132,   9.01716){\circle*{0.2}} 
\put(   49.12843,   8.31005){\circle*{0.2}} 
\put(   45.71421,  16.55269){\circle*{0.2}} 
\put(   44.30000,  16.55269){\circle*{0.2}} 
\put(   48.42132,   6.60294){\circle*{0.2}} 
\put(   39.17868,   6.60294){\circle*{0.2}} 
\put(   38.47157,   8.31005){\circle*{0.2}} 
\put(   40.17868,   5.60294){\circle*{0.2}} 
\put(   40.88579,   5.89584){\circle*{0.2}} 
\put(   38.47157,   9.72426){\circle*{0.2}} 
\put(   37.47157,  10.72426){\circle*{0.2}} 
\put(   37.76447,  10.01716){\circle*{0.2}} 
\put(   38.47157,  11.72426){\circle*{0.2}} 
\put(   37.76447,  11.43137){\circle*{0.2}} 
\put(   38.47157,  13.13848){\circle*{0.2}} 
\put(   39.47157,  14.13848){\circle*{0.2}} 
\put(   41.88579,   4.89584){\circle*{0.2}} 
\put(   43.30000,   4.89584){\circle*{0.2}} 
\put(   40.88579,  15.55269){\circle*{0.2}} 
\put(   39.17868,  14.84558){\circle*{0.2}} 
\put(   40.17868,  15.84558){\circle*{0.2}} 
\put(   43.30000,  16.55269){\circle*{0.2}} 
\put(   41.88579,  16.55269){\circle*{0.2}} 
\put(   43.59289,   4.18873){\circle*{0.2}} 
\put(   45.00711,   4.18873){\circle*{0.2}} 
\put(   46.71421,   4.89584){\circle*{0.2}} 
\put(   47.71421,   5.89584){\circle*{0.2}} 
\put(   49.42132,  14.84558){\circle*{0.2}} 
\put(   49.12843,  15.55269){\circle*{0.2}} 
\put(   48.42132,  15.84558){\circle*{0.2}} 
\put(   49.12843,  14.13848){\circle*{0.2}} 
\put(   50.12843,  13.13848){\circle*{0.2}} 
\put(   50.12843,  11.72426){\circle*{0.2}} 
\put(   46.71421,  16.55269){\circle*{0.2}} 
\put(   50.83553,  11.43137){\circle*{0.2}} 
\put(   50.12843,   9.72426){\circle*{0.2}} 
\put(   50.83553,  10.01716){\circle*{0.2}} 
\put(   50.12843,   8.31005){\circle*{0.2}} 
\put(   49.12843,   7.31005){\circle*{0.2}} 
\put(   45.00711,  17.25980){\circle*{0.2}} 
\put(   44.30000,  17.55269){\circle*{0.2}} 
\put(   43.59289,  17.25980){\circle*{0.2}} 
\put(   49.42132,   6.60294){\circle*{0.2}} 
\put(   48.42132,   5.60294){\circle*{0.2}} 
\put(   49.12843,   5.89584){\circle*{0.2}} 
\put(   38.47157,   5.89584){\circle*{0.2}} 
\put(   39.17868,   5.60294){\circle*{0.2}} 
\put(   38.47157,   7.31005){\circle*{0.2}} 
\put(   37.47157,   8.31005){\circle*{0.2}} 
\put(   37.76447,   9.01716){\circle*{0.2}} 
\put(   39.47157,   4.89584){\circle*{0.2}} 
\put(   40.88579,   4.89584){\circle*{0.2}} 
\put(   36.76447,  10.01716){\circle*{0.2}} 
\put(   36.76447,  11.43137){\circle*{0.2}} 
\put(   37.76447,  12.43137){\circle*{0.2}} 
\put(   37.47157,  13.13848){\circle*{0.2}} 
\put(   38.47157,  14.13848){\circle*{0.2}} 
\put(   42.59289,   4.18873){\circle*{0.2}} 
\put(   40.88579,  16.55269){\circle*{0.2}} 
\put(   39.17868,  15.84558){\circle*{0.2}} 
\put(   38.47157,  15.55269){\circle*{0.2}} 
\put(   39.47157,  16.55269){\circle*{0.2}} 
\put(   42.59289,  17.25980){\circle*{0.2}} 
\put(   43.59289,   3.18873){\circle*{0.2}} 
\put(   44.30000,   3.48162){\circle*{0.2}} 
\put(   46.00711,   4.18873){\circle*{0.2}} 
\put(   45.00711,   3.18873){\circle*{0.2}} 
\put(   47.71421,   4.89584){\circle*{0.2}} 
\put(   47.42132,   4.18873){\circle*{0.2}} 
\put(   50.12843,  15.55269){\circle*{0.2}} 
\put(   50.12843,  14.13848){\circle*{0.2}} 
\put(   49.12843,  16.55269){\circle*{0.2}} 
\put(   47.71421,  16.55269){\circle*{0.2}} 
\put(   50.83553,  12.43137){\circle*{0.2}} 
\put(   47.42132,  17.25980){\circle*{0.2}} 
\put(   46.71421,  17.55269){\circle*{0.2}} 
\put(   46.00711,  17.25980){\circle*{0.2}} 
\put(   51.83553,  11.43137){\circle*{0.2}} 
\put(   51.54264,  10.72426){\circle*{0.2}} 
\put(   50.83553,   9.01716){\circle*{0.2}} 
\put(   51.83553,  10.01716){\circle*{0.2}} 
\put(   50.12843,   7.31005){\circle*{0.2}} 
\put(   50.83553,   7.60294){\circle*{0.2}} 
\put(   45.00711,  18.25980){\circle*{0.2}} 
\put(   43.59289,  18.25980){\circle*{0.2}} 
\put(   50.12843,   5.89584){\circle*{0.2}} 
\put(   49.12843,   4.89584){\circle*{0.2}} 
\put(   37.47157,   5.89584){\circle*{0.2}} 
\put(   38.47157,   4.89584){\circle*{0.2}} 
\put(   37.76447,   6.60294){\circle*{0.2}} 
\put(   37.47157,   7.31005){\circle*{0.2}} 
\put(   36.76447,   7.60294){\circle*{0.2}} 
\put(   36.76447,   9.01716){\circle*{0.2}} 
\put(   40.17868,   4.18873){\circle*{0.2}} 
\put(   41.59289,   4.18873){\circle*{0.2}} 
\put(   36.05736,  10.72426){\circle*{0.2}} 
\put(   36.76447,  12.43137){\circle*{0.2}} 
\put(   37.47157,  14.13848){\circle*{0.2}} 
\put(   37.76447,  14.84558){\circle*{0.2}} 
\put(   42.59289,   3.18873){\circle*{0.2}} 
\put(   41.59289,  17.25980){\circle*{0.2}} 
\put(   40.88579,  17.55269){\circle*{0.2}} 
\put(   40.17868,  17.25980){\circle*{0.2}} 
\put(   38.47157,  16.55269){\circle*{0.2}} 
\put(   42.59289,  18.25980){\circle*{0.2}} 
\put(   44.30000,   2.48162){\circle*{0.2}} 
\put(   46.00711,   3.18873){\circle*{0.2}} 
\put(   46.71421,   3.48162){\circle*{0.2}} 
\put(   45.71421,   2.48162){\circle*{0.2}} 
\put(   48.42132,   4.18873){\circle*{0.2}} 
\put(   47.42132,   3.18873){\circle*{0.2}} 
\put(   50.12843,  16.55269){\circle*{0.2}} 
\put(   50.83553,  14.84558){\circle*{0.2}} 
\put(   50.83553,  13.43137){\circle*{0.2}} 
\put(   49.12843,  17.55269){\circle*{0.2}} 
\put(   48.42132,  17.25980){\circle*{0.2}} 
\put(   51.83553,  12.43137){\circle*{0.2}} 
\put(   51.54264,  13.13848){\circle*{0.2}} 
\put(   47.42132,  18.25980){\circle*{0.2}} 
\put(   46.00711,  18.25980){\circle*{0.2}} 
\put(   52.54264,  10.72426){\circle*{0.2}} 
\put(   51.83553,   9.01716){\circle*{0.2}} 
\put(   51.54264,   8.31005){\circle*{0.2}} 
\put(   50.83553,   6.60294){\circle*{0.2}} 
\put(   45.71421,  18.96690){\circle*{0.2}} 
\put(   44.30000,  18.96690){\circle*{0.2}} 
\put(   50.12843,   4.89584){\circle*{0.2}} 
\put(   37.47157,   4.89584){\circle*{0.2}} 
\put(   36.76447,   6.60294){\circle*{0.2}} 
\put(   37.76447,   4.18873){\circle*{0.2}} 
\put(   39.17868,   4.18873){\circle*{0.2}} 
\put(   36.05736,   8.31005){\circle*{0.2}} 
\put(   35.76447,   9.01716){\circle*{0.2}} 
\put(   36.05736,   9.72426){\circle*{0.2}} 
\put(   40.17868,   3.18873){\circle*{0.2}} 
\put(   40.88579,   3.48162){\circle*{0.2}} 
\put(   41.59289,   3.18873){\circle*{0.2}} 
\put(   35.05736,  10.72426){\circle*{0.2}} 
\put(   36.05736,  11.72426){\circle*{0.2}} 
\put(   35.76447,  12.43137){\circle*{0.2}} 
\put(   36.76447,  13.43137){\circle*{0.2}} 
\put(   36.05736,  13.13848){\circle*{0.2}} 
\put(   36.76447,  14.84558){\circle*{0.2}} 
\put(   37.76447,  15.84558){\circle*{0.2}} 
\put(   41.88579,   2.48162){\circle*{0.2}} 
\put(   43.30000,   2.48162){\circle*{0.2}} 
\put(   41.59289,  18.25980){\circle*{0.2}} 
\put(   40.17868,  18.25980){\circle*{0.2}} 
\put(   39.17868,  17.25980){\circle*{0.2}} 
\put(   37.47157,  16.55269){\circle*{0.2}} 
\put(   38.47157,  17.55269){\circle*{0.2}} 
\put(   37.76447,  17.25980){\circle*{0.2}} 
\put(   43.30000,  18.96690){\circle*{0.2}} 
\put(   41.88579,  18.96690){\circle*{0.2}} 
\put(   44.30000,   1.48162){\circle*{0.2}} 
\put(   45.00711,   1.77452){\circle*{0.2}} 
\put(   46.71421,   2.48162){\circle*{0.2}} 
\put(   49.42132,   4.18873){\circle*{0.2}} 
\put(   48.42132,   3.18873){\circle*{0.2}} 
\put(   49.12843,   3.48162){\circle*{0.2}} 
\put(   50.83553,  17.25980){\circle*{0.2}} 
\put(   50.12843,  17.55269){\circle*{0.2}} 
\put(   50.83553,  15.84558){\circle*{0.2}} 
\put(   51.83553,  14.84558){\circle*{0.2}} 
\put(   51.54264,  14.13848){\circle*{0.2}} 
\put(   48.42132,  18.25980){\circle*{0.2}} 
\put(   52.54264,  13.13848){\circle*{0.2}} 
\put(   52.54264,  11.72426){\circle*{0.2}} 
\put(   46.71421,  18.96690){\circle*{0.2}} 
\put(   53.54264,  10.72426){\circle*{0.2}} 
\put(   53.24975,  11.43137){\circle*{0.2}} 
\put(   52.54264,   9.72426){\circle*{0.2}} 
\put(   53.24975,  10.01716){\circle*{0.2}} 
\put(   52.54264,   8.31005){\circle*{0.2}} 
\put(   51.54264,   7.31005){\circle*{0.2}} 
\put(   51.83553,   6.60294){\circle*{0.2}} 
\put(   50.83553,   5.60294){\circle*{0.2}} 
\put(   51.54264,   5.89584){\circle*{0.2}} 
\put(   45.00711,  19.67401){\circle*{0.2}} 
\put(   44.30000,  19.96690){\circle*{0.2}} 
\put(   50.83553,   4.18873){\circle*{0.2}} 
\put(   36.76447,   4.18873){\circle*{0.2}} 
\put(   36.76447,   5.60294){\circle*{0.2}} 
\put(   35.76447,   6.60294){\circle*{0.2}} 
\put(   36.05736,   5.89584){\circle*{0.2}} 
\put(   36.05736,   7.31005){\circle*{0.2}} 
\put(   37.76447,   3.18873){\circle*{0.2}} 
\put(   38.47157,   3.48162){\circle*{0.2}} 
\put(   39.17868,   3.18873){\circle*{0.2}} 
\put(   35.05736,   8.31005){\circle*{0.2}} 
\put(   35.05736,   9.72426){\circle*{0.2}} 
\put(   39.47157,   2.48162){\circle*{0.2}} 
\put(   40.88579,   2.48162){\circle*{0.2}} 
\put(   34.35025,  10.01716){\circle*{0.2}} 
\put(   35.05736,  11.72426){\circle*{0.2}} 
\put(   34.35025,  11.43137){\circle*{0.2}} 
\put(   35.05736,  13.13848){\circle*{0.2}} 
\put(   36.05736,  14.13848){\circle*{0.2}} 
\put(   35.76447,  14.84558){\circle*{0.2}} 
\put(   36.76447,  15.84558){\circle*{0.2}} 
\put(   36.05736,  15.55269){\circle*{0.2}} 
\put(   41.88579,   1.48162){\circle*{0.2}} 
\put(   42.59289,   1.77452){\circle*{0.2}} 
\put(   43.30000,   1.48162){\circle*{0.2}} 
\put(   40.88579,  18.96690){\circle*{0.2}} 
\put(   39.17868,  18.25980){\circle*{0.2}} 
\put(   39.47157,  18.96690){\circle*{0.2}} 
\put(   36.76447,  17.25980){\circle*{0.2}} 
\put(   37.76447,  18.25980){\circle*{0.2}} 
\put(   43.30000,  19.96690){\circle*{0.2}} 
\put(   42.59289,  19.67401){\circle*{0.2}} 
\put(   41.88579,  19.96690){\circle*{0.2}} 
\put(   43.59289,    .77452){\circle*{0.2}} 
\put(   45.00711,    .77452){\circle*{0.2}} 
\put(   29.47619,  10.00000){\circle*{0.2}} 
\put(   27.47619,  10.00000){\circle*{0.2}} 
\put(   29.18329,  10.70711){\circle*{0.2}} 
\put(   27.76908,   9.29289){\circle*{0.2}} 
\put(   28.47619,  11.00000){\circle*{0.2}} 
\put(   28.47619,   9.00000){\circle*{0.2}} 
\put(   27.76908,  10.70711){\circle*{0.2}} 
\put(   29.18329,   9.29289){\circle*{0.2}} 
\put(   28.47619,  10.00000){\circle*{0.2}} 
\end{picture}		   
\caption{{\it Left:} The strip ${\bf S}_{1,3}$ and the window ${\bf W}_{1,3}$ in the case of a
1D physical space ${\bf E}_1$ embedded into a three-dimensional superspace.
{\it Centre:} A one-shell $C_8$-cluster $\mathcal{C}_2$. 
{\it Right:} A fragment of a set $\Omega $ defined by $\mathcal{C}_2$.
The nearest neighbours of any point $q$ of $\Omega $ belong to $q+\mathcal{C}_2$, 
which is a copy of $\mathcal{C}_2$ with the center at point $q$.} 
\end{figure}

Consider the {\it strip} corresponding to ${\bf W}_{2,k}$ (see figure 1)
\begin{equation} 
{\bf S}_{2,k}=\{ x\in \mathbb{R}^{k}\ |\ \pi _2^\perp x\in {\bf W}_{2,k} \ \} 
\end{equation}
and define for each $(i_1,i_2,i_3)\in \mathcal{I}_{2,k}$ the number
\begin{equation} d_{i_1i_2i_3}=\max_{\alpha _j \in \{ -1/2,\, 1/2\}}
\left| 
\begin{array}{ccc}
\alpha _1& \alpha _2 & \alpha _3 \\
v_{1i_1} & v_{1i_2} & v_{1i_3}\\
v_{2i_1} & v_{2i_2} & v_{2i_3}
\end{array} \right| .
\end{equation}
A point $x\in \mathbb{R}^{k}$ 
belongs to the strip ${\bf S}_{2,k}$ if and only if 
\begin{equation} -d_{i_1i_2i_3}\leq 
\left| %
		   
\caption{{\it Left:} A one-shell $C_{10}$-cluster and a fragment of the corresponding 
quasiperiodic set. 
{\it Right:} A fragment of the quasiperiodic set defined by a two-shell $C_{10}$-cluster,
obtained by using strip projection method in a ten-dimensional superspace. 
The starting cluster is a covering cluster, 
but for most of the points the occupation is extremely low.} 
\end{figure}

The set defined in terms of the strip projection method \cite{E,KKL,K,KN}
\begin{equation} 
\Omega =\pi _2({\bf S}_{2,k}\cap \mathbb{Z}^{k})=
\{ \ \pi _2x\ | \ \ x\in {\bf S}_{2,k}\cap \mathbb{Z}^{k}\ \} 
\end{equation}
can be regarded as a packing of translated partially occupied
copies of $\mathcal{C}_2$. Since
\begin{equation} 
\pi _2e_i=(\langle e_i,w_1\rangle , \langle e_i, w_2\rangle )
=(v_{1i},v_{2i})=v_i 
\end{equation} 
we get
\begin{eqnarray}
 \pi _2&(\{ &x\pm e_1,\, x\pm e_2,\, ...,\, x\pm e_{k} \}\cap {\bf S}_{2,k})\nonumber \\
 &\subseteq &\{ \pi _2x\pm v_1,\, \pi _2x\pm v_2\, ...,\, 
\pi _2x\pm v_{k}\}=\pi _2x+\mathcal{C}_2 
\end{eqnarray}
that is, the neighbours of any point
$\pi _2x\!\in \!\Omega $ belong to the translated copy 
$\pi _2x\!+\!\mathcal{C}_2$ of $\mathcal{C}_2 $.

A larger class of aperiodic pattens can be obtained by translating the strip ${\bf S}_{2,k}$. 
For each $t\!\in \!\mathbb{R}^k$ the set
\begin{equation} 
\Omega =\pi _2((t+{\bf S}_{2,k})\cap \mathbb{Z}^{k})=
\{ \ \pi _2x\ | \ \ x\!-\!t\in {\bf S}_{2,k}\ \ {\rm and}\ \ x\in \mathbb{Z}^{k}\ \} 
\end{equation}
is a packings of partially occupied copies of $\mathcal{C}_2$. Some particular examples can be
seen in figures 1-3.

\section{Modified strip projection method}

The algorithm based on the strip projection method presented in the previous section is very efficient. Hundreds of
points of $\Omega $ can be obtained in only a few minutes for rather complicated $G$-clusters
$\mathcal{C}_2$, but, as one can remark in figures 1-3, for most of the points of $\Omega $ 
the occupation of the corresponding cluster is very low. On the other hand, the images concerning
the quasicrystal structure obtained by high resolution transmission electron microscopy show 
the presence of a significant percentage of fully occupied clusters. 

The number $n(x)$ of the neighbours of a point $\pi _2x\!\in \!\Omega $ occurring in $\Omega $ 
corresponds to the number of the points of the set $\{ x\pm e_1,\, x\pm e_2,\, ...,\, x\pm e_{k} \}$ 
belonging to the strip $t+{\bf S}_{2,k}$.
In our modified strip projection method we firstly determine $n(x)$ for all the points of $\mathbb{Z}^k$
lying in the fragment of the strip we intend to project, and then we project the points in the decreasing 
order of the occupation number $n(x)$.
In the case when $n(x)$ represents more than $p\%$ of all the points of the cluster $\mathcal{C}_2$ we project all the 
arithmetic neighbours of $x$ (lying inside or outside the strip $t+{\bf S}_{2,k}$).
We choose $p$ such that to avoid the superposition of the fully occupied clusters. 
The projection of a point $x$ with $n(x)$ less than $p\%$ of all the points of the cluster $\mathcal{C}_2$ is added to the 
pattern only if it is not too close to the already obtained points. We get in this way a discrete set 
$\tilde {\Omega }$ containing fully occupied copies of the cluster $\mathcal{C}_2$. 

In the structure analysis of quasicrystals, the experimental diffraction image is compared with the diffraction 
image of the mathematical model, regarded as a set of scatterers. In order to compute the diffraction image 
of a discrete set $\Omega $ one has to use the Fourier transform and to identify $\Omega $  with the Dirac comb
$\sum_{\omega \in \Omega }\delta _\omega $, which will also be denoted by $\Omega $ .
The set $\Omega $ defined in terms of the strip projection method is an infinite set, but the set 
$\tilde {\Omega }_0$ we can effectively generate is evidently a finite set, obtained by starting from
a finite fragment $\Omega _0$ of $\Omega $. This is not very bad since any quasicrystal has a finite number of atoms.
Nevertheless, a fragment of $\Omega $ or $\tilde {\Omega }$ can not be an acceptable model for a quasicrystal 
unless it is large enough.

\begin{figure}			
\setlength{\unitlength}{1.5mm}	   

\caption{{\it Left:} A set ${\Omega }_0$ containing 923 points 
defined by starting from a $C_{12}$-cluster $\mathcal{C}_2$. 
{\it Centre:} The cluster $\mathcal{C}_2$. 
{\it Right:} For the diffraction pattern ${\Omega _0^*}$ of ${\Omega _0}$ please see [5].}  		    
\end{figure}

\begin{figure}			
\setlength{\unitlength}{1.5mm}	   
%
\caption{{\it Left:} The set $\tilde{\Omega }_0$ containing 1019 points 
corresponding to the set $\Omega _0$ from the previous figure, defined by using the modified strip 
projection method. {\it Centre:} The cluster $\mathcal{C}_2$.
{\it Right:} For the diffraction pattern $\tilde{\Omega _0^*}$ of $\tilde{\Omega _0}$  please see [5].}  
\end{figure}

The diffraction pattern corresponding to $\Omega _0$ is related to the function
\begin{equation}
\Omega _0^*:\mathbb{R}^2\longrightarrow [0, \infty)\qquad  
\Omega _0^*(\xi )=\left|\mathcal{F}[\Omega _0](\xi )\right|^2
=\left|\sum_{\omega \in \Omega _0}{\rm e}^{{\rm i}\langle \omega , \xi \rangle }\right|^2
\end{equation}
where $\mathcal{F}[\Omega _0]$ means the Fourier transform of the distribution 
$\Omega_0=\sum_{\omega \in \Omega _0}\delta _\omega $.
In figure 3 we present a fragment $\Omega _0$ of the set $\Omega $ corresponding to a $C_{12}$-cluster and 
the set $\left\{ \xi \in \mathbb{R}^2\ |\ \Omega _0^*(\xi )> \frac{1}{1000}\Omega _0^*(0) \right\}$ in 
order to illustrate the shape and symmetry properties of the diffraction image of $\Omega _0$. 
The case of the set $\tilde{\Omega _0}$ corresponding to $\Omega _0$, obtained by using the modified strip 
projection method in $\mathbb{R}^6$, is presented in figure 4. 

It is an open problem if the diffraction properties of the sets obtained by using the modified strip 
projection method are similar to those of sets obtained by the non-modified version. 
A mathematical answer seems to be difficult, but some suggestions in this direction can be obtained by analysing
larger fragments. By using our computer program the fragment containing 923 points presented in figure 3 
and its diffraction pattern are obtained in one minute. In two hours one can obtain about 16000 points.
The fragment generated by the modified method presented in figure 4 can be obtained in two minutes.

\section{Quasiperiodic packings of icosahedral clusters}

The icosahedral group $Y=235$ can be defined in terms of generators and relations as 
\begin{equation} Y=\langle a,b\ |\ a^5=b^2=(ab)^3=e \rangle \end{equation}
and the rotations $a,\, b :\mathbb{R}^3\longrightarrow \mathbb{R}^3$
\begin{equation}\label{Y} \fl \begin{array}{l}
a(\alpha ,\beta  ,\gamma )=
\left(\frac{\tau -1}{2}\alpha -\frac{\tau }{2}\beta  +\frac{1}{2}\gamma ,
\ \frac{\tau }{2}\alpha +\frac{1}{2}\beta  +\frac{\tau -1}{2}\gamma ,
\ -\frac{1}{2}\alpha +\frac{\tau -1}{2}\beta  
+\frac{\tau }{2}\gamma \right)\\[1mm]
b(\alpha ,\beta  ,\gamma )=(-\alpha ,-\beta  ,\gamma ).
\end{array} \end{equation}
where $\tau =(1+\sqrt{5})/2$, generate  an irreducible representation of $Y$ in 
$\mathbb{R}^3$.
In the case of this representation there are the trivial orbit 
$Y(0,0,0)=\{ (0,0,0)\}$ of length 1, the orbits
\begin{equation}
Y(\alpha ,\alpha \tau ,0)=\{ g(\alpha ,\alpha \tau ,0)\ |\ g\in Y\}\qquad 
{\rm where}\quad \alpha \in (0,\infty )
\end{equation}
of length 12 (vertices of a regular icosahedron), the orbits
\begin{equation}
Y(\alpha ,\alpha ,\alpha )=\{ g(\alpha ,\alpha ,\alpha )\ |\ g\in Y\}\qquad 
{\rm where}\quad \alpha \in (0,\infty )
\end{equation}
of length 20 (vertices of a regular dodecahedron), the orbits
\begin{equation}
Y(\alpha ,0,0)=\{ g(\alpha ,0,0)\ |\ g\in Y\}\qquad 
{\rm where}\quad \alpha \in (0,\infty )
\end{equation}
of length 30 (vertices of an icosidodecahedron), and all the other orbits are 
of length 60. 

If the set symmetric with respect to the origin
\begin{equation}
\mathcal{C}_3=\{ v_1,\, v_2,\, ...,\, v_k,\, -v_1,\, -v_2,\, ...,\, -v_k \}
\end{equation}
where $v_1=(v_{11},v_{21},v_{31})$, ..., 
$v_k=(v_{1k},v_{2k},v_{3k})$, is a finite union of orbits of $Y$ then the vectors 
\begin{equation} \begin{array}{l}
w_1=(v_{11},v_{12},...,v_{1k})\\
w_2=(v_{21},v_{22},...,v_{2k})\\
w_3=(v_{31},v_{32},...,v_{3k})
\end{array}
\end{equation}
from $\mathbb{R}^{k}$ are orthogonal 
\begin{equation} 
\langle w_1,w_2\rangle =\langle w_2,w_3\rangle =\langle w_3,w_1\rangle =0
\end{equation}
and have the same norm $\kappa =||w_1||=||w_2||=||w_3||$.
They allow us to identify the physical space with the three-dimensional subspace
\begin{equation} 
{\bf E}_3=\{ \ \alpha w_1+\beta w_2+\gamma w_3 \ | \ \alpha,\, \beta ,\, \gamma \in \mathbb{R}\ \} 
\end{equation}
of the superspace $\mathbb{R}^k$. 
The orthogonal projection on ${\bf E}_3$ of a vector $x\in \mathbb{R}^k$ is the vector
\begin{equation} 
\pi _3\, x= \left\langle x,\frac{w_1}{\kappa }\right\rangle\frac{w_1}{\kappa }
             +\left\langle x,\frac{w_2}{\kappa }\right\rangle\frac{w_2}{\kappa}
             +\left\langle x,\frac{w_3}{\kappa }\right\rangle\frac{w_3}{\kappa}
\end{equation}
The orthogonal projector corresponding to the orthogonal complement 
\begin{equation} 
{\bf E}_3^\perp =\{ \ x\in \mathbb{R}^k\ |\ 
\langle x,y\rangle =0\ {\rm for\ all}\ y\in {\bf E}_3\ \} 
\end{equation}
is $\pi _3^\perp :\mathbb{R}^{k}\longrightarrow {\bf E}_3^\perp$, \, 
$\pi _3^\perp x=x-\pi _3\, x$.
If we describe ${\bf E}_3$ by using the orthogonal basis 
$\{ \kappa ^{-2}w_1,\, \kappa ^{-2}w_2,\, \kappa ^{-2}w_3\}$ then the expression 
in coordinates of $\pi_3$ is
\begin{equation} 
\pi _3: \mathbb{R}^{k}\longrightarrow \mathbb{R}^3\qquad 
\pi _3x=(\langle x,w_1\rangle , \langle x,w_2\rangle , \langle x,w_3\rangle ). 
\end{equation}

The projection ${\bf W}_{3,k}=\pi _3^\perp ({\Lambda }_{k})$ of the unit hypercube
${\Lambda }_{k}$   is a polyhedron in the $(k\!-\!3)$-dimensional subspace ${\bf E}_3^\perp $, 
and each $(k\!-\!4)$-dimensional face of ${\bf W}_{3,k}$ is the projection of a 
$(k\!-\!4)$-dimensional face of ${\Lambda }_{k}$.
Each $(k\!-\!4)$-face of ${\Lambda }_{k}$ 
is parallel to $(k\!-\!4)$ of the vectors $e_1$, $e_2$, ..., $e_k$ and orthogonal to four of them.
There exist sixteen $(k\!-\!4)$-faces of ${\Lambda }_{k}$ 
orthogonal to the distinct vectors $e_{i_1}$, $e_{i_2}$, $e_{i_3}$, $e_{i_4}$, and the set 
\begin{equation}\left\{ \ x=(x_1,x_2,...,x_{k})\ \left| \ \begin{array}{lcl}
x_i\in \{ -1/2,\, 1/2\} & {\rm if}&   i\in \{ i_1,\, i_2,\, i_3,\, i_4\} \\
x_i=0 & {\rm if}& i\not\in \{ i_1,\, i_2,\, i_3,\, i_4\}
\end{array} \right. \right\} \end{equation}
contains one and only one point from each of them.
There are 
\begin{equation} \left( \begin{array}{c}
k\\
4
\end{array}\right) =\frac{k(k-1)(k-2)(k-3)}{24} \end{equation}
sets of $2^4$ parallel $(k\!-\!4)$-faces of ${\Lambda }_{k}$, and we label them by using the 
elements of the set
\begin{equation} \fl
\mathcal{I}_{3,k}=
\left\{ (i_1,i_2,i_3,i_4)\in \mathbb{Z}^4\ \left|\ 
            \begin{array}{rl}
            1\leq i_1\leq k-3,\ \ \ & i_1+1\leq i_2\leq k-2,\\ 
            i_2+1\leq i_3\leq k-1,\ \ \  & i_3+1\leq i_4\leq k
            \end{array} \right.
\right\}. 
\end{equation}
A point $x=(x_1,x_2,...,x_k)\in \mathbb{R}^k$ belongs to the
strip ${\bf S}_{3,k}$ if and only if 
\begin{equation} -d_{i_1i_2i_3i_4}\leq 
\left| \begin{array}{cccc}
x_{i_1}& x_{i_2} & x_{i_3} & x_{i_4}\\
v_{1i_1} & v_{1i_2} & v_{1i_3} & v_{1i_4}\\
v_{2i_1} & v_{2i_2} & v_{2i_3} & v_{2i_4}\\
v_{3i_1} & v_{3i_2} & v_{3i_3} & v_{3i_4}
\end{array} \right| 
\leq d_{i_1i_2i_3i_4}
\end{equation}
for each $(i_1,i_2,i_3,i_4)\in \mathcal{I}_{3,k}$, where
\begin{equation} d_{i_1i_2i_3i_4}=\max_{\alpha _j \in \{ -1/2,\, 1/2\}}
\left| \begin{array}{cccc}
\alpha _1& \alpha _2 & \alpha _3 & \alpha _4 \\
v_{1i_1} & v_{1i_2} & v_{1i_3} & v_{1i_4}\\
v_{2i_1} & v_{2i_2} & v_{2i_3} & v_{2i_4}\\
v_{3i_1} & v_{3i_2} & v_{3i_3} & v_{3i_4}
\end{array} \right|.\end{equation}

For each $t\!\in \!\mathbb{R}^k$, the pattern defined in terms of the strip projection method \cite{E,KKL,K,KN}
\begin{equation} 
\Omega =\pi _3((t+{\bf S}_{3,k})\cap \mathbb{Z}^k)=
\{ \pi _3x\ |\ \ x\!-\!t\in {\bf S}_{3,k}\ \ {\rm and}\ \ x\in \mathbb{Z}^k\ \} 
\end{equation}
can be regarded as a quasiperiodic packing of copies of the starting cluster 
$\mathcal{C}_3$.

The algorithm based on the strip projection method presented above is very efficient. 
In the case of a three-shell $Y$-cluster formed by the vertices of a regular icosahedron, 
a regular dodecahedron and an icosidodecahedron we use a 31-dimensional superspace, 
${\bf W}_{3,k}$ is a polyhedron lying in the 28-dimensional subspace ${\bf E}_3^\perp $
bounded by 31465 pairs of parallel 27-dimensional faces,  but we obtain 400-500 points in less than
10 minutes \cite{C2}. With the modification indicated in the previous section we can obtain
quasiperiodic packings of multi-shell icosahedral clusters containing a significant 
percentage of fully occupied clusters.

\section{Concluding remarks}

Some of the most remarkable tilings and discrete quasiperiodic sets used in 
quasicrystal physics are obtained by using strip projection 
method in a superspace of dimension four, five or six, and the 
projection of a unit hypercube as a window of selection \cite{C4,E,K}.
The mathematical results presented above allow one to use this very 
elegant method in superspaces of dimension much higher, and to generate 
discrete quasiperiodic sets with a more complicated structure by starting from the 
symmetry group $G$ and the local structure described by a covering cluster $\mathcal{C}$. 
In our approach the window (which, generally, is a polyhedron with hundreds or thousands faces)
is described in a simple way and we have to compute only determinants of order three or four, 
independently of the dimension of the superspace we use. These mathematical results have allowed 
us to obtain some very efficient computer programs for our algorithm \cite{C5}. 
Hundreds of points of our mathematical models can be obtained in only a few minutes.

The quasiperiodic set generated by starting from a $G$-cluster $\mathcal{C}$ is a packing of 
partially occupied copies of $\mathcal{C}$, but for most of these copies the occupation is very low.
The main purpose of the paper is to present a modified version of the strip projection method. We project
certain points lying outside the strip and do not project certain points lying inside the strip in 
order to favour the apperance of fully occupied clusters. More exactly, we start from the pattern
generated by the standard strip projection method and help the clusters with occupation above a
certain threshold ($p\%$) to complete their configuration up to a fully occupied clusters.
We project a minimum number of points lying outside the strip, and avoid to project certain points
lying inside the strip. 

The symmetry group $G$ corresponding to a real quasicrystal can be deduced from the diffraction images, 
and the covering cluster $\mathcal{C}$ can be chosen by analyzing the real structure information obtained
by high-resolution transmission electron microscopy. Our modified strip projection method allows one
to generate a mathematical model, and to compute the corresponding diffraction image by using the 
Fourier transform. If the agreement with the experimental data is not acceptable one has to look for
a more suitable covering cluster $\mathcal{C}$.

\newpage

\section*{The computer program in FORTRAN 90 used in the case of figure 3}

\begin{verbatim}
! PLEASE INDICATE HOW MANY POINTS DO YOU WANT TO ANALYSE 
      INTEGER, PARAMETER :: N = 6000 

! PLEASE INDICATE THE DIMENSION  M  OF THE SUPERSPACE
      INTEGER, PARAMETER :: M = 6

! PLEASE INDICATE THE RADIUS OF THE PATTERN
      REAL, PARAMETER :: R = 9.0

	  INTEGER I, J, K, L, I1, I2, I3, JJ, J1, J2
	  REAL D1, D2, D3, PR	  
        REAL, DIMENSION(M) :: V, W, TRANSLATION, WJ, EPSILON
    	  INTEGER, DIMENSION(N) :: CLUSTER		  
	  REAL, DIMENSION(2,M) :: BASIS			
	  REAL, DIMENSION(2,2) :: C12 
	  REAL, DIMENSION(1:M-2,2:M-1,3:M) :: STRIP
	  REAL, DIMENSION(N,M) :: POINTS, STRIPOINTS
	  REAL, DIMENSION(N + M) :: XPOINT, YPOINT
	  REAL, DIMENSION(200,200) :: FOURIER
	  REAL, DIMENSION(2,40000) :: PATTERN
	  COMPLEX II
	  II=(0,1)

	  EPSILON = 0.0001
	    
! PLEASE INDICATE THE COORDINATES OF A POINT OF THE CLUSTER
      BASIS(1,1) = 1.0   
	  BASIS(2,1) = 0.0  

! PLEASE INDICATE THE TRANSLATION OF THE STRIP YOU WANT TO USE	  
	  TRANSLATION = 0.1 
	                      
	  C12(1,1) = SQRT(3.0) / 2.0	 
	  C12(1,2) = -1.0 / 2.0
	  C12(2,1) = 1.0 / 2.0
	  C12(2,2) = SQRT(3.0) / 2.0
	  DO J = 2, 6
	   DO I = 1, 2
        BASIS(I,J) = C12(I,1) * BASIS(1,J-1) &
		           + C12(I,2) * BASIS(2,J-1)
      END DO	 
      END DO	
      STRIP=0 
      DO I1 =1, M-2		
	  DO I2 =I1+1, M-1	  
	  DO I3 =I2+1, M	  
	    DO D1 =-0.5, 0.5  
	    DO D2 =-0.5, 0.5  
	    DO D3 =-0.5, 0.5
	    PR = D1 * BASIS(1,I2) * BASIS(2,I3) + &
		     D3 * BASIS(1,I1) * BASIS(2,I2) + &
		     D2 * BASIS(1,I3) * BASIS(2,I1) - &
			 D3 * BASIS(1,I2) * BASIS(2,I1) - &
		     D1 * BASIS(1,I3) * BASIS(2,I2) - &
			 D2 * BASIS(1,I1) * BASIS(2,I3)
	    IF ( PR > STRIP(I1,I2,I3) ) STRIP(I1,I2,I3) = PR
	    END DO
	    END DO	  
		END DO	 
		IF( STRIP(I1,I2,I3) .EQ. 0 ) STRIP(I1,I2,I3)=N * SUM( BASIS(1,:) ** 2)
	  END DO	  
	  END DO	 
	  END DO	 
	  PRINT*, 'COORDINATES OF THE POINTS OF THE ONE-SHELL C12-CLUSTER:'
	  DO J = 1, M
	  PRINT*, J, BASIS(1,J), BASIS(2,J)
	  END DO
	  PRINT*, '* STRIP TRANSLATED BY THE VECTOR WITH COORDINATES:'
	  PRINT*,    TRANSLATION
	  PRINT*, '* PLEASE WAIT A FEW MINUTES OR MORE,&
	             DEPENDING ON THE NUMBER OF ANALYSED POINTS'        
   	  POINTS = 0
	  STRIPOINTS = 0
	  POINTS(1,:) = ANINT( TRANSLATION)
	  STRIPOINTS(1,:) = ANINT( TRANSLATION)
	  K = 1			        
	  L = 0				
      DO I = 1, N
   	  V = POINTS(I, : )	- TRANSLATION
      JJ = 1				
      DO I1 =1, M-2			
	   DO I2 =I1+1, M-1		
	    DO I3 =I2+1, M
	    PR = V(I1) * BASIS(1,I2) * BASIS(2,I3) + &
		     V(I3) * BASIS(1,I1) * BASIS(2,I2) + &
	         V(I2) * BASIS(1,I3) * BASIS(2,I1) - &
			 V(I3) * BASIS(1,I2) * BASIS(2,I1) - &
		     V(I1) * BASIS(1,I3) * BASIS(2,I2) - &
			 V(I2) * BASIS(1,I1) * BASIS(2,I3)
         IF ( ABS(PR) > STRIP(I1,I2,I3) ) JJ = 0
		END DO			
	   END DO			
	  END DO	 	
    IF( JJ .EQ. 1 ) THEN
	  I3 = 1
	  DO J = 1, L
	  WJ = ABS(POINTS(I,:) - STRIPOINTS(J,:))
	  IF( ALL(WJ < EPSILON) ) I3 = 0
	  END DO
	  IF( I3 == 1 .AND. SUM( V * V) < R **2 ) THEN	 
      L = L + 1							  
      STRIPOINTS(L,:) = POINTS(I,:)
	  ELSE
	  END IF  
      DO I1 = 1, M		  
	   DO I2 = -1, 1, 2		  
	     W = POINTS(I, : )				 
	     W(I1) = W(I1) + I2
	     I3 = 0									
	      DO J = 1, K
		  WJ = ABS(W - POINTS(J,:))
		 IF( ALL( WJ < EPSILON)) I3 = 1	 
	      END DO								 
	       IF ( I3 == 0 .AND. K < N ) THEN	  
	       K = K + 1					
	       POINTS(K, : ) = W				
	       ELSE							
	       END IF
	    END DO						   
	   END DO						   
	   ELSE							 
	   END IF						 
      END DO
      CLUSTER = 0
      DO I = 1, L
      DO J1 = 1, M
        DO J2 = -1, 1, 2
		W = STRIPOINTS(I,:)	- TRANSLATION
        W(J1) = W(J1) + J2
      JJ = 1				
      DO I1 =1, M-2			
	   DO I2 =I1+1, M-1		
	    DO I3 =I2+1, M
	    PR = W(I1) * BASIS(1,I2) * BASIS(2,I3) + &
		     W(I3) * BASIS(1,I1) * BASIS(2,I2) + &
	         W(I2) * BASIS(1,I3) * BASIS(2,I1) - &
			 W(I3) * BASIS(1,I2) * BASIS(2,I1) - &
		     W(I1) * BASIS(1,I3) * BASIS(2,I2) - &
			 W(I2) * BASIS(1,I1) * BASIS(2,I3)
         IF ( ABS(PR) > STRIP(I1,I2,I3) ) JJ = 0
		END DO			
	   END DO			
	  END DO	 	
      IF( JJ .EQ. 1 ) CLUSTER(I) = CLUSTER(I) + 1   
	  END DO
      END DO
      END DO
	  DO J = 1, L
	  XPOINT(J) = SUM( STRIPOINTS(J,:) * BASIS(1,:) )  
      YPOINT(J) = SUM( STRIPOINTS(J,:) * BASIS(2,:) )
	  END DO
	  PRINT*, 'NUMBER OF ANALYSED POINTS :', K
      PRINT*, 'NUMBER OF OBTAINED POINTS :', L
	   DO I = 1, 100
	   DO J = 1, 100
	     D1=0.0
	     DO I1 = 1, L
	     D1=D1+EXP( II * (-1.5+I*0.03)*XPOINT(I1)+ &
		            II * (-1.5+J*0.03)*YPOINT(I1) )
		 END DO
		 FOURIER(I,J)=(ABS(D1))**2
	   END DO
	   END DO
	   I2=0
	   DO I = 1, 100
	   DO J = 1, 100
	     IF(FOURIER(I,J)>0.001*L**2) THEN
		 I2=I2+1
		 PATTERN(1,I2)=-1.5+I*0.03
		 PATTERN(2,I2)=-1.5+J*0.03
		 ELSE
		 END IF
	   END DO
	   END DO
	 PRINT*, 'INDICATE THE NAME OF A FILE WITH EXTENSION tex FOR RESULTS'
	 WRITE(4,60)					  
 60  FORMAT('\documentclass{article}	 &
              \begin{document}			 &
              \begin{figure}			&
              \setlength{\unitlength}{1.5mm}	  &
              \begin{picture}(50,20)(0,0) '  &
			  '\put(32.0,20.0){\circle*{0.2}} ')
	  DO J = 1, L 
	  IF( CLUSTER(J) < M+1 ) THEN
	  WRITE(4,65) 10+XPOINT(J), 20+YPOINT(J)	
 65   FORMAT( '\put( 'F10.5','F10.5,'){\circle*{0.2}} ')
     ELSE
	  END IF
	  END DO
	   DO J = 1, L 
	  IF( CLUSTER(J) > M ) THEN
	  WRITE(4,70) 10+XPOINT(J), 20+YPOINT(J)	
 70   FORMAT( '\put( 'F10.5','F10.5,'){\circle{0.4}} ')
     ELSE
	  END IF
	  END DO
 	  DO J = 1, 6 
	  WRITE(4,72) 32+BASIS(1,J), 20+BASIS(2,J)	
 72   FORMAT( '\put( 'F10.5','F10.5,'){\circle*{0.2}} ')
	  WRITE(4,73) 32-BASIS(1,J), 20-BASIS(2,J)	
 73   FORMAT( '\put( 'F10.5','F10.5,'){\circle*{0.2}} ')
	  END DO
  	 WRITE(4,75)					  
 75  FORMAT(' \setlength{\unitlength}{1.8mm}')
 	  DO J = 1, I2 
	  WRITE(4,80) 45+10*PATTERN(1,J), 17+10*PATTERN(2,J)	
 80   FORMAT( '\put( 'F10.5','F10.5,'){\circle*{0.1}} ')
	  END DO
 	  WRITE(4,90) 
 90   FORMAT( '\end{picture}		   &
               \caption{Quasiperiodic set obtained by using &
                 the strip projection method } &
               \end{figure}		&
              \end{document}')
	  PRINT*, '* COMPILE THE OBTAINED FILE AND SEE THE  ".dvi" FILE' 
	  END
\end{verbatim}
\newpage

\section*{The computer program in FORTRAN 90 used in the case of figure 4}

\begin{verbatim}
! PLEASE INDICATE HOW MANY POINTS DO YOU WANT TO ANALYSE 
      INTEGER, PARAMETER :: N = 6000 

! PLEASE INDICATE THE DIMENSION  M  OF THE SUPERSPACE
      INTEGER, PARAMETER :: M = 6

! PLEASE INDICATE THE RADIUS OF THE PATTERN
      REAL, PARAMETER :: R = 9.0

	  INTEGER I, J, K, L, I1, I2, I3, JJ, J1, J2, L1
	  REAL D1, D2, D3, PR, XP, YP	  
        REAL, DIMENSION(M) :: V, W, TRANSLATION, WJ, EPSILON
    	  INTEGER, DIMENSION(N) :: CLUSTER		  
	  REAL, DIMENSION(2,M) :: BASIS			
	  REAL, DIMENSION(2,2) :: C12 
	  REAL, DIMENSION(1:M-2,2:M-1,3:M) :: STRIP
	  REAL, DIMENSION(N,M) :: POINTS, STRIPOINTS
	  REAL, DIMENSION(N + M) :: XPOINT, YPOINT
	  REAL, DIMENSION(200,200) :: FOURIER
	  REAL, DIMENSION(2,40000) :: PATTERN
	  COMPLEX II
	  II=(0,1)

	  EPSILON = 0.0001
	    
! PLEASE INDICATE THE COORDINATES OF A POINT OF THE CLUSTER
      BASIS(1,1) = 1.0   
	  BASIS(2,1) = 0.0  

! PLEASE INDICATE THE TRANSLATION OF THE STRIP YOU WANT TO USE	  
	  TRANSLATION = 0.1 
	                      
	  C12(1,1) = SQRT(3.0) / 2.0	 
	  C12(1,2) = -1.0 / 2.0
	  C12(2,1) = 1.0 / 2.0
	  C12(2,2) = SQRT(3.0) / 2.0
	  DO J = 2, 6
	   DO I = 1, 2
        BASIS(I,J) = C12(I,1) * BASIS(1,J-1) &
		           + C12(I,2) * BASIS(2,J-1)
	!    BASIS(I,6+J) = C12(I,1) * BASIS(1,5+J) &
	!                 + C12(I,2) * BASIS(2,5+J)
       END DO	 
      END DO	
      STRIP=0 
      DO I1 =1, M-2		
	  DO I2 =I1+1, M-1	  
	  DO I3 =I2+1, M	  
	    DO D1 =-0.5, 0.5  
	    DO D2 =-0.5, 0.5  
	    DO D3 =-0.5, 0.5
	    PR = D1 * BASIS(1,I2) * BASIS(2,I3) + &
		     D3 * BASIS(1,I1) * BASIS(2,I2) + &
		     D2 * BASIS(1,I3) * BASIS(2,I1) - &
			 D3 * BASIS(1,I2) * BASIS(2,I1) - &
		     D1 * BASIS(1,I3) * BASIS(2,I2) - &
			 D2 * BASIS(1,I1) * BASIS(2,I3)
	    IF ( PR > STRIP(I1,I2,I3) ) STRIP(I1,I2,I3) = PR
	    END DO
	    END DO	  
		END DO	 
		IF( STRIP(I1,I2,I3) .EQ. 0 ) STRIP(I1,I2,I3)=N * SUM( BASIS(1,:) ** 2)
	  END DO	  
	  END DO	 
	  END DO	 
	  PRINT*, 'COORDINATES OF THE POINTS OF THE ONE-SHELL C12-CLUSTER:'
	  DO J = 1, M
	  PRINT*, J, BASIS(1,J), BASIS(2,J)
	  END DO
	  PRINT*, '* STRIP TRANSLATED BY THE VECTOR WITH COORDINATES:'
	  PRINT*,    TRANSLATION
	  PRINT*, '* PLEASE WAIT A FEW MINUTES OR MORE,&
	             DEPENDING ON THE NUMBER OF ANALYSED POINTS'        
   	  POINTS = 0
	  STRIPOINTS = 0
	  POINTS(1,:) = ANINT( TRANSLATION)
	  STRIPOINTS(1,:) = ANINT( TRANSLATION)
	  K = 1			        
	  L = 0				
      DO I = 1, N
   	  V = POINTS(I, : )	- TRANSLATION
      JJ = 1				
      DO I1 =1, M-2			
	   DO I2 =I1+1, M-1		
	    DO I3 =I2+1, M
	    PR = V(I1) * BASIS(1,I2) * BASIS(2,I3) + &
		     V(I3) * BASIS(1,I1) * BASIS(2,I2) + &
	         V(I2) * BASIS(1,I3) * BASIS(2,I1) - &
			 V(I3) * BASIS(1,I2) * BASIS(2,I1) - &
		     V(I1) * BASIS(1,I3) * BASIS(2,I2) - &
			 V(I2) * BASIS(1,I1) * BASIS(2,I3)
         IF ( ABS(PR) > STRIP(I1,I2,I3) ) JJ = 0
		END DO			
	   END DO			
	  END DO	 	
    IF( JJ .EQ. 1 ) THEN
	  I3 = 1
	  DO J = 1, L
	  WJ = ABS(POINTS(I,:) - STRIPOINTS(J,:))
	  IF( ALL(WJ < EPSILON) ) I3 = 0
	  END DO
	  IF( I3 == 1 .AND. SUM( V * V) < R **2 ) THEN	 
      L = L + 1							  
      STRIPOINTS(L,:) = POINTS(I,:)
	  ELSE
	  END IF  
      DO I1 = 1, M		  
	   DO I2 = -1, 1, 2		  
	     W = POINTS(I, : )				 
	     W(I1) = W(I1) + I2
	     I3 = 0									
	      DO J = 1, K
		  WJ = ABS(W - POINTS(J,:))
		 IF( ALL( WJ < EPSILON)) I3 = 1	 
	      END DO								 
	       IF ( I3 == 0 .AND. K < N ) THEN	  
	       K = K + 1					
	       POINTS(K, : ) = W				
	       ELSE							
	       END IF
	    END DO						   
	   END DO						   
	   ELSE							 
	   END IF						 
      END DO
      CLUSTER = 0
      DO I = 1, L
      DO J1 = 1, M
        DO J2 = -1, 1, 2
		W = STRIPOINTS(I,:)	- TRANSLATION
        W(J1) = W(J1) + J2
      JJ = 1				
      DO I1 =1, M-2			
	   DO I2 =I1+1, M-1		
	    DO I3 =I2+1, M
	    PR = W(I1) * BASIS(1,I2) * BASIS(2,I3) + &
		     W(I3) * BASIS(1,I1) * BASIS(2,I2) + &
	         W(I2) * BASIS(1,I3) * BASIS(2,I1) - &
			 W(I3) * BASIS(1,I2) * BASIS(2,I1) - &
		     W(I1) * BASIS(1,I3) * BASIS(2,I2) - &
			 W(I2) * BASIS(1,I1) * BASIS(2,I3)
         IF ( ABS(PR) > STRIP(I1,I2,I3) ) JJ = 0
		END DO			
	   END DO			
	  END DO	 	
      IF( JJ .EQ. 1 ) CLUSTER(I) = CLUSTER(I) + 1   
	  END DO
      END DO
      END DO
	  I1 = L
	  DO I = 1, I1
	  IF ( CLUSTER(I) > M) THEN
	  DO J1 = 1, M
	  DO J2 = -1, 1, 2
	  W = STRIPOINTS(I,:)
	  W(J1) = W(J1) + J2
	  	   I3 = 0									
	      DO J = 1, L
		  WJ = ABS(W - STRIPOINTS(J,:))
		 IF( ALL( WJ < EPSILON)) I3 = 1	 
	      END DO								 
	       IF ( I3 == 0 ) THEN	  
	       L = L + 1					
	       STRIPOINTS(L, : ) = W				
	       ELSE							
	       END IF
	   END DO
	   END DO
	   ELSE
	   END IF
	   END DO
	   L1=0
	  DO I = 1, I1
	  IF ( CLUSTER(I) > M) THEN
	  L1=L1+1
	  XPOINT(L1) = SUM( STRIPOINTS(I,:) * BASIS(1,:) )  
      YPOINT(L1) = SUM( STRIPOINTS(I,:) * BASIS(2,:) )
	  ELSE							
	  END IF
	  END DO
	  DO J = I1+1, L
	  L1=L1+1
	  XPOINT(L1) = SUM( STRIPOINTS(J,:) * BASIS(1,:) )  
      YPOINT(L1) = SUM( STRIPOINTS(J,:) * BASIS(2,:) )
	  END DO
	  D1=4.0
	  DO I =2, M
	  IF(((BASIS(1,1)+BASIS(1,I))**2 + (BASIS(2,1)+BASIS(2,I))**2 ) < D1 ) &
			D1=(BASIS(1,1)+BASIS(1,I))**2 + (BASIS(2,1)+BASIS(2,I))**2
	  END DO
	  DO I =2, M
	  IF(((BASIS(1,1)-BASIS(1,I))**2 + (BASIS(2,1)-BASIS(2,I))**2 ) < D1 ) &
			D1=(BASIS(1,1)-BASIS(1,I))**2 + (BASIS(2,1)-BASIS(2,I))**2
	  END DO
	  DO I = 1, I1
	  IF ( CLUSTER(I) <= M) THEN
	  D2=4.0
	  XP = SUM( STRIPOINTS(I,:) * BASIS(1,:) )  
      YP = SUM( STRIPOINTS(I,:) * BASIS(2,:) )
	  DO J = 1, L1
	  IF( ((XP - XPOINT(J))**2 + (YP - YPOINT(J))**2 ) <  D2 )  &
			 D2=(XP - XPOINT(J))**2 + (YP - YPOINT(J))**2 
	  END DO
	  IF( D2 > 0.9*D1 ) THEN
	  L1=L1+1
	  XPOINT(L1)=XP
	  YPOINT(L1)=YP
	  ELSE							
	  END IF
	  
	  ELSE							
	  END IF
	  END DO

	   DO I = 1, 100
	   DO J = 1, 100
	     D1=0.0
	     DO I1 = 1, L1
	     D1=D1+EXP( II * (-1.5+I*0.03)*XPOINT(I1)+ &
		            II * (-1.5+J*0.03)*YPOINT(I1) )
		 END DO
		 FOURIER(I,J)=(ABS(D1))**2
	   END DO
	   END DO
	   I2=0
	   DO I = 1, 100
	   DO J = 1, 100
	     IF(FOURIER(I,J)>0.0015*L**2) THEN
		 I2=I2+1
		 PATTERN(1,I2)=-1.5+I*0.03
		 PATTERN(2,I2)=-1.5+J*0.03
		 ELSE
		 END IF
	   END DO
	   END DO
	   PRINT*, 'NUMBER OF ANALYSED POINTS :', K
      PRINT*, 'NUMBER OF OBTAINED POINTS :', L1

	 PRINT*, 'INDICATE THE NAME OF A FILE WITH EXTENSION tex FOR RESULTS'
	 WRITE(4,60)					  
 60  FORMAT('\documentclass{article}	 &
              \begin{document}			 &
              \begin{figure}			&
              \setlength{\unitlength}{1.5mm}	  &
              \begin{picture}(50,20)(0,0) '		 &
			  '\put(32.0,20.0){\circle*{0.2}} ')
	  DO J = 1, L1 
	  IF( CLUSTER(J) < M+1 ) THEN
	  WRITE(4,65) 10+XPOINT(J), 20+YPOINT(J)	
 65   FORMAT( '\put( 'F10.5','F10.5,'){\circle*{0.2}} ')
     ELSE
	  END IF
	  END DO
	   DO J = 1, L1 
	  IF( CLUSTER(J) > M ) THEN
	  WRITE(4,70) 10+XPOINT(J), 20+YPOINT(J)	
 70   FORMAT( '\put( 'F10.5','F10.5,'){\circle*{0.2}} ')
     ELSE
	  END IF
	  END DO
 	  DO J = 1, 6 
	  WRITE(4,72) 32+BASIS(1,J), 20+BASIS(2,J)	
 72   FORMAT( '\put( 'F10.5','F10.5,'){\circle*{0.2}} ')
	  WRITE(4,73) 32-BASIS(1,J), 20-BASIS(2,J)	
 73   FORMAT( '\put( 'F10.5','F10.5,'){\circle*{0.2}} ')
	  END DO
  	 WRITE(4,75)					  
 75  FORMAT(' \setlength{\unitlength}{1.8mm}')
 	  DO J = 1, I2 
	  WRITE(4,80) 45+10*PATTERN(1,J), 17+10*PATTERN(2,J)	
 80   FORMAT( '\put( 'F10.5','F10.5,'){\circle*{0.1}} ')
	  END DO
 	  WRITE(4,90) 
 90   FORMAT( '\end{picture}		   &
                 \caption{Quasiperiodic set obtained by using &
                 the modified strip projection method } &
               \end{figure}		&
              \end{document}')
	  PRINT*, '* COMPILE THE OBTAINED FILE AND SEE THE  ".dvi" FILE' 
	  END
\end{verbatim}
\newpage

\section*{Acknowledgment}

This  research was supported by the grants CERES 4-129 and CEx05-D11-03.

\end{document}